\documentstyle[12pt,draft,epsf]{article}

\newcommand \be  {\begin{equation}}
\newcommand \bea {\begin{eqnarray} \nonumber }
\newcommand \ee  {\end{equation}}
\newcommand \eea {\end{eqnarray}}

\begin{document}

\title{ON TOY AGING\\[1.0em]}

\author{
Enzo MARINARI$^{(a),(b)}$ and Giorgio PARISI$^{(c)}$\\[.5em]
$(a)$
 NPAC and Physics Department,\\
 Syracuse University,\\
 Syracuse, N.Y. 13244, U.S.A\\
$(b)$
 Dipartimento di Fisica and INFN,\\
 Universit\`a di Roma {\em Tor Vergata},\\
 Via della Ricerca Scientifica 2,\\
 00133 Roma, Italy\\
$(c)$
 Dipartimento di Fisica and INFN,\\
 Universit\`a di Roma {\em La Sapienza},\\
 P. A. Moro 2,\\
 00185 Roma, Italy\\[0.5em]
 {\footnotesize \tt MARINARI@ROMA1.INFN.IT PARISI@ROMA1.INFN.IT}\\[1.0em]
}


\maketitle

\begin{abstract}

We consider the dynamics of a simple one dimensional model and we
discuss the phenomenon of {\em aging} (i.e., the strong dependence of the
dynamical correlation functions over the waiting time).  Our model is
the so-called {\em random random walk,} the toy model of a directed
polymer evolving in a random medium.

\end{abstract}

\vfill
{\hfill {\bf \footnotesize ROMA LA SAPIENZA 962-93, cond-mat/9308003} }
\vfill

\newpage

Aging is a very interesting phenomenon that has been observed in spin
glasses \cite{AGESWE,AGEUSA,AGEFRA,AGETWO}, but it likely to be
present in many other materials (for example in rubber), and to be a
crucial signature of the behavior of a strongly disordered system.
In brief we can say we have {\em aging} if the response of the system
to a perturbation strongly depends on the time $t_w$ (waiting time)
during which the system has been kept in the low temperature phase
before starting the measurement.

More precisely in an aging experiment one brings the system in the low
temperature phase at time $t_0=0$. One leaves the system quiescent up
to the time $t_w$, and at $t_w$ she adds a small perturbation (small
enough that the linear response theory can be applied). One eventually
measures the response function $R(t,t_w)$ at time $t+t_w$.
Alternatively one can measure the correlation function $C(t,t_w)$
among two quantities measured at time $t_w$ and at time $t+t_w$.

In the region where the observation time $t$ is very smaller than the
waiting time $t_w$, $t \ll t_w$, $R$ and $C$ are independent from
$t_w$ and they are related by the fluctuation-dissipation theorem.
Aging is present if $R$ and $C$ are strongly dependent on $t_w$ in the
region where $t$ is of the order of $t_w$.  The phenomenon of aging
has been observed non only in real spin glasses, but also in numerical
simulations of short range \cite{RIEGER} and long range \cite{CUKURI}
spin glasses. In this last case the mean field approach
\cite{MEPAVI,PARBO2} is exact, so that an analytic study of aging
should be possible.

It seems very natural that in the region of large times a simple
scaling behavior could hold. Different suggestions have been put
forward.  One simple possibility \cite{BOU} is that in the region
where both $t$ and $t_w$ are large

\be
  C(t,t_w)= t^{-\mu}f(\frac{t}{t_w})\ ,
  \label{E_ANS}
\ee
where the exponent $\mu$ may be equal to zero.  Unfortunately
at the moment we do not have a complete justification of this
formula.

Waving our hands we can sketch a simple qualitative explanation. We
suppose that the system is described by a potential with a corrugated
landscape, where many barriers of all possible heights are present. Such a
system (considered in the thermodynamic limit) is always out of
equilibrium. Its dynamics is given by the never ending search of the
absolute minimum by crossing potential barriers that can be arbitrary
large.

Let us assume that the available phase space does not increase too
much with the height of the barriers that have been crossed.  In this
case we can suppose that at time $t_w$ and temperature $T$ the system
has explored all the phase space that can be reached from the origin
by crossing barriers that are smaller than $T \ln(t_w)$.  The system
therefore will remain near the bottom of the explored phase space up
to the moment at which it crosses one other large barrier. This can
happen only at times of the order of $t_w$.  If we assume that the
shape of the deepest minimum found does not depend on the size of the
explored region we are lead to eq. (\ref{E_ANS}). In this way we can
explain the dependence on $\frac{t}{t_w}$. One would need a more
careful analysis to derive the value of $\mu$, which may depend on the
detailed quantity whose correlation function is computed. It should be
noted that if

\be
  \lim_{t \to \infty} \lim_{t_W \to \infty}C(t,t_w)= {\rm const} \neq 0
\ee
(i.e. if $C$ goes to a non zero finite limit in the region where both $t$ and
$t_w$ are large and $t \ll t_w$) then $\mu=0$.

The aim of this note is to discuss the behavior of a very simple
system, the one dimensional random walk in random environment
\cite{MPRW}.  In this simple case many analytic results are available
(for a review see \cite{BOUREV}). Our numerical simulations will show
that the aging phenomenon is present also in such a simple setting.

The model we consider is a particle which equilibrium probability
distribution is proportional to

\be
  P(x) \propto e^{-\beta (V (x)+\lambda x^2)}\ ,
  \label{E_EQU}
\ee
where $V$ is a random Gaussian quantity with zero average and
correlations

\be
  \overline{(V(x)-V(y))^2} = |x-y|^\alpha\ .
\ee

In the case $\alpha=1$ (the one that we will mainly consider in the
following) the force $F(x) \equiv -\frac{dV}{dx}$ is uncorrelated from
point to point. One introduce the contribution $\lambda x^2$ to
regularize the static equilibrium distribution, and $\lambda \to 0$ in
the relevant limit.  At equilibrium this model coincide with the toy
model that was introduced in the study of the behavior of interfaces
in random medium (see \cite{MEZPAR} and references therein).

It is known that \cite{PARISI}

\bea
  \overline{\langle x\rangle^2} &\sim& \lambda^{-4/3}\ ,\\ \nonumber
  \overline{\langle x^2\rangle-\langle x\rangle^2 } &\sim& \lambda^{-1}\ ,\\
  \overline{\ln(\langle x^2\rangle-\langle x\rangle^2 )} &\sim& O(1).
\eea

The first equation implies that for non-zero $\lambda$ the particle is
contained inside a region of radius of order $\lambda^{-2/3}$. In the
$\lambda \to 0$ limit the particle is always at infinity at
equilibrium (the potential $V$ is unbound).

In most cases the particle is localized around the minimum. The most
probable value of $\langle x^2 \rangle - \langle x\rangle^2$ is of order
one, although rare events, in which the potential has two widely
separated nearly degenerate minima, do contribute strongly to the
average of $\langle x^2 \rangle - \langle x \rangle^2$, making it of
order $\lambda^{-1}$.

The dynamics of this model is a random walk biased to produce the
equilibrium distribution (\ref{E_EQU}) at large times.  More precisely
the model can be defined by introducing a one dimensional lattice. If
the particle is at point $x$ at time $t$ ($x$ and $t$ taking integer
values) then $p_x$ is the probability that at time $t+1$ the particle
is at site $x+1$, and $1-p_x$ the probability that it is at site $x-1$.
$p_x$ is a random variable with zero average (typically uniformly
distributed in $(0,1)$, not including the interval limits).

It is easy to see that the equilibrium potential corresponding to this
dynamics satisfy the condition

\be
  V(x+1)-V(x)= -\ln(\frac{p_x}{1-p_{x+1}})\ .
  \label{E_POT}
\ee
Consequently at large distance the difference of the potential
$V(x)-V(y)$ are approximately Gaussian, with $\alpha=1$.

This dynamics has been widely investigated \cite{BOUREV}. It can be
proven \cite{SINAI} that the particle to go from $x$ to $y$
has to cross (with high probability) a barrier that is of order
$|x-y|^{1/2}$. If the system is in $x=0$ at $t=0$, at a large time
$t$ it will be in $x \sim \ln(t)^2$.

In the following we will discuss the phenomenon of aging in this
model. The simplest quantity one could study is the correlation
function

\be
  C(t,t_w)= \overline{(x(t+t_w)-x(t_w))^2}\ ,
  \label{E_CX2}
\ee
that for large $t$ (in the asymptotic region studied by Sinai, where
$t\gg t_w$) behaves as $\ln(t)^4$.  But this is not a good choice. In
the region of large times, for $t \ll t_w$, this function behaves as
$\ln(t)^3$. The reasons for this behavior are quite clear.  If we
assume that the probability distribution for the dynamics at time $t$
may be mimed by the probability distribution for the statics, where
$\lambda$ is chosen so that the value of $\langle x\rangle$ coincides
with the correct one, we find from the previous equations that
$\langle x^2 \rangle-\langle x\rangle^2$ is of order $\ln(t)^3$.  In
other words the system has explored a region of size $\ln(t)^2$ and
the probability of having two nearly equal minima inside this region
vanishes as $\frac{1}{\ln(t)}$ as follows from the analysis of ref.
\cite{PARISI}.

The previous analysis implies that the aging properties of the
function $C$ are not simple and the scaling law in eq. (\ref{E_ANS})
cannot hold with $\mu=0$. So we found convenient to
consider the following correlation

\be
  L(t,t_w) \equiv \overline{\ln( (x(t+t_w)-x(t_w))^2 )}\ ,
\ee
or equivalently $M(t,t_w)\equiv\exp(L(t,t_w))$. $M$ is not far from the
most likely value of $(x(t+t_w)-x(t_w))^2$. In this case we expect that
in the region of large times, when $t \ll t_w$ ,

\be
  M(t,t_w) = {\rm constant} + O(\ln(t)^{-1})\ .
\ee

All that said, it seemed natural to set up
a numerical simulation to verify the validity of the scaling law:

\be
  M(t,t_w) = f(\frac{t}{t_w}) + O(\ln(t)^{-1})\ .
\ee

We have generated $15\times 10^{3}$ realizations of the
one-dimensional random potential. For each of these realizations of
the random potential we have observed the random walker travelling for
$2^K$ steps and we have measured the correlation $M(t,t_w)$ at times
$t$ and $t_w$ equal to $2^k$, with $k=1,...,K-1$. The spatial lattice
had an infinite extent, i.e. the particle would never hit a boundary.
We have taken $K=21$, i.e., performed order of $2 \times 10^6$ sweeps
per sample.

We show in Fig.~1 $\langle \ln(|x(t)|) \rangle$ versus $\ln(\ln(t+1))$
for different waiting time $t_w$, scaled logarithmically. In Fig.~1 we
have drawn continuous lines joining the points of equal $t_w$. Higher
curves correspond to lower $t_w$. The first continuous line from the
top joins points taken after no waiting ($t_w=0$), the second the ones
with $t_w=1$, the third the ones with $t_w=2$, the fourth the ones
with $t_w=4$ and so on, with an exponential progression.

\begin{figure}
\epsffile[22 206 565 690]{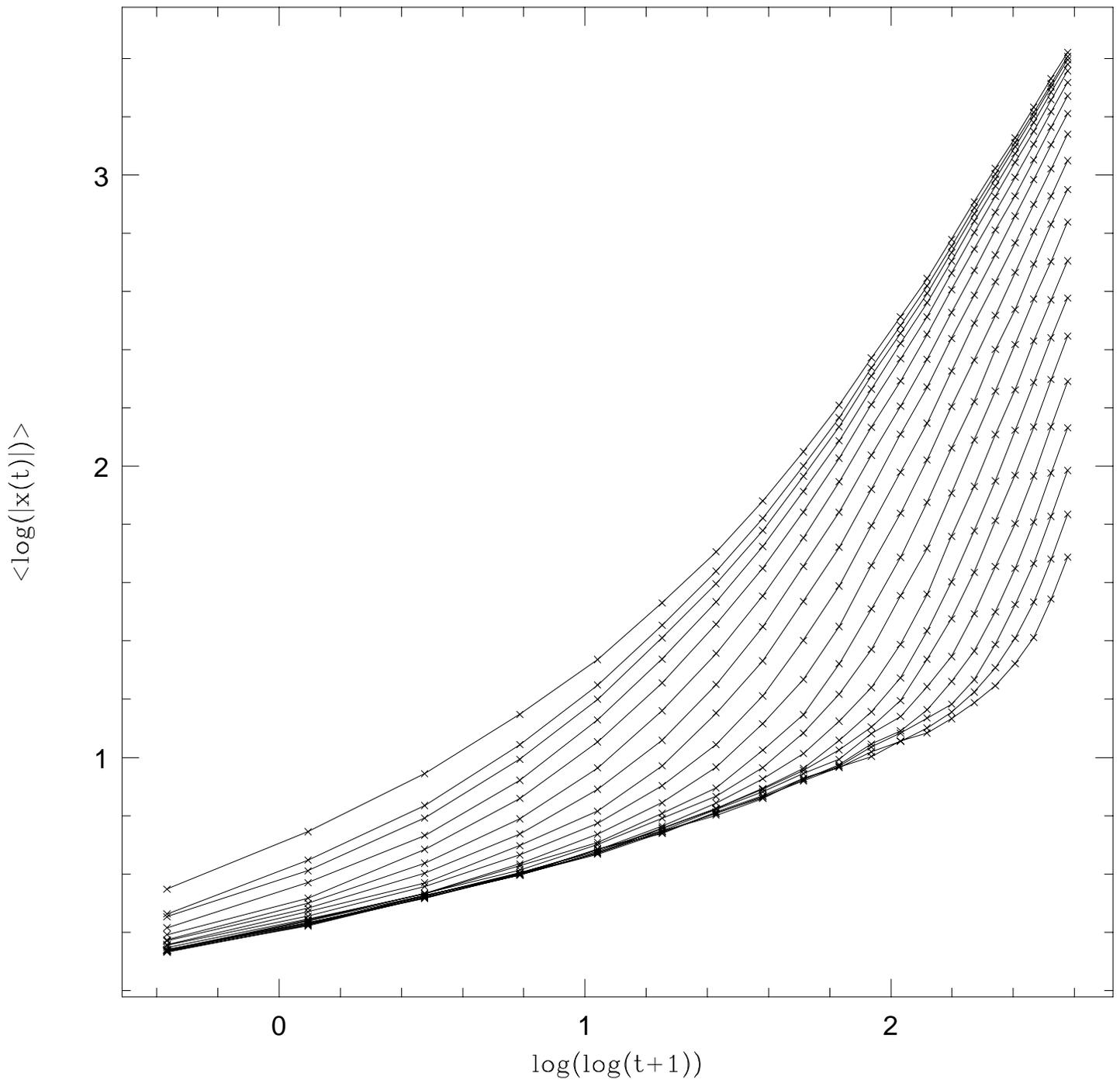}
\caption[a]{The average logarithm of the distance reached by the
walker after $t$ steps as a function of $\ln(\ln(t))$. Here the
lines join points with constant $t_w$. Lines from top to
bottom are from lower to higher $t_w$.}
\end{figure}

The upper lines of Fig.~1 have $t \gg t_w$, and in the right part of
the plot they scale with good precision as expected. Such straight
lines have indeed a slope very close to $2$.

In Fig.~2 we plot the same points, but we join points of {\em
constant} $\frac{t}{t_w}$ {\em ratio.} Here the $\frac{t}{t_w}$
scaling is quite clear. Indeed the lines at intermediate values of
$\frac{t}{t_w}$ tend for large $t$ to a constant value.  The distance
reached from the walker at time $t$ only depends over the ratio
$\frac{t}{t_w}$. If we double the observation time $t$ but we also
double the waiting time $t_w$ the average distance covered by the
walker does not change.  Figures $1$ and $2$, and the other figures we
will show, allow us to draw the main conclusion of this note. {\em For
large $t$ and $t_w$, $t$ of order $t_w$, the correlation functions
behave as an universal function of $\frac{t}{t_w}$.} The simplicity of
the $1d$ (toy) model allowed us to get a numerical very compelling
evidence for such an effect.

\begin{figure}
\epsffile[22 206 565 690]{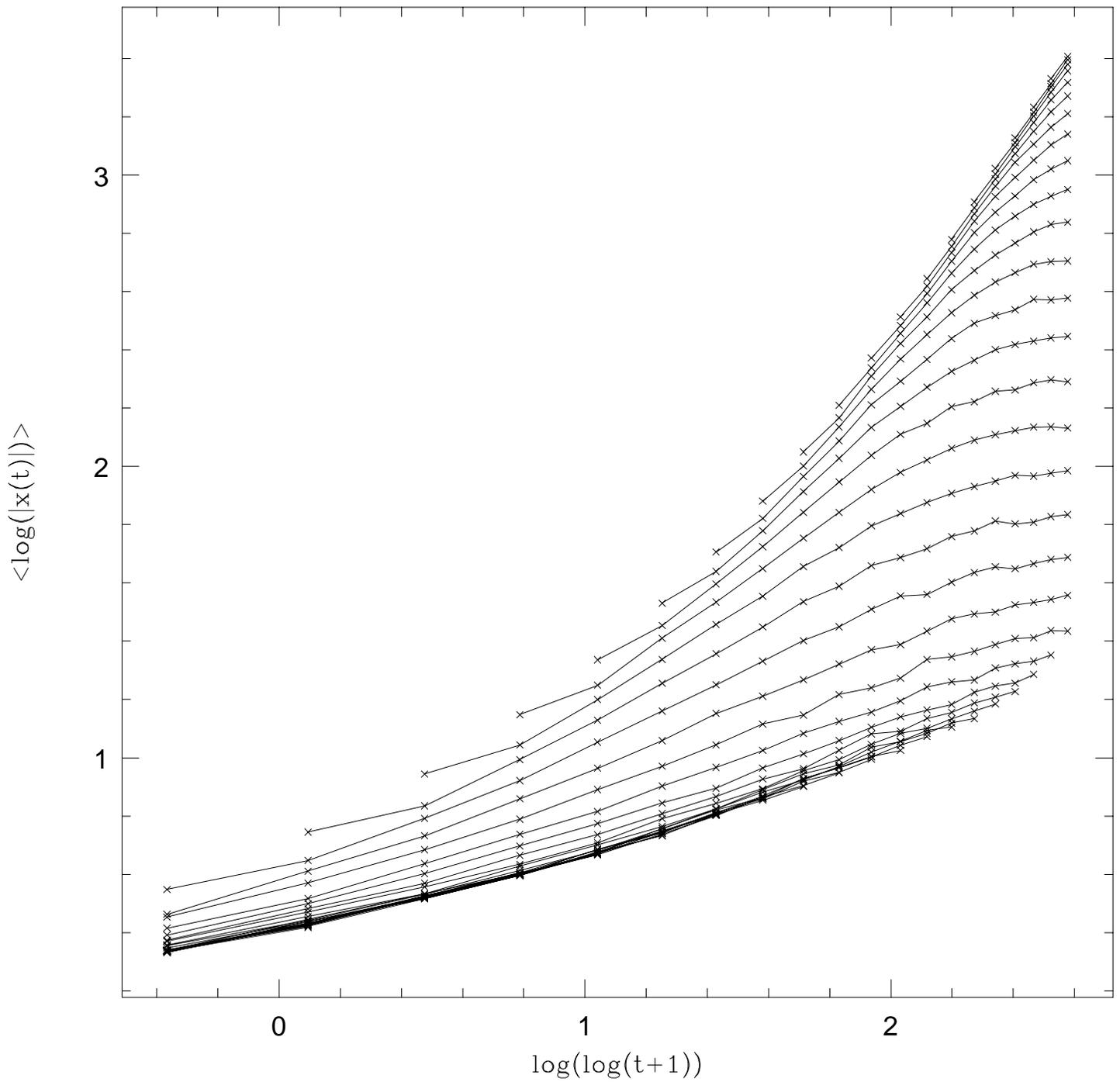}
\caption[a]{Same points of Fig.~1, but here the lines join points
with constant $\frac{t}{t_w}$.}
\end{figure}

In Fig.~3 we plot again the same data, but this time we select the
ones with both $t$ and $t_w$ larger than $32$, and we plot
$\exp(\langle \ln(|x(t)|) \rangle)$ versus $\ln(\frac{t+1}{t_w+1})$.
The data fall with good precision on a single scaling curve. In the
left part of the $x$-axis (small $t$ as compared to $t_w$) the
universal curve is a constant (aside from small corrections), while in
the large $\frac{t}{t_w}$ region it behaves as $\ln(\frac{t}{t_w})^2$
(according to the Sinai \cite{SINAI} scaling law).

\begin{figure}
\epsffile[22 206 565 690]{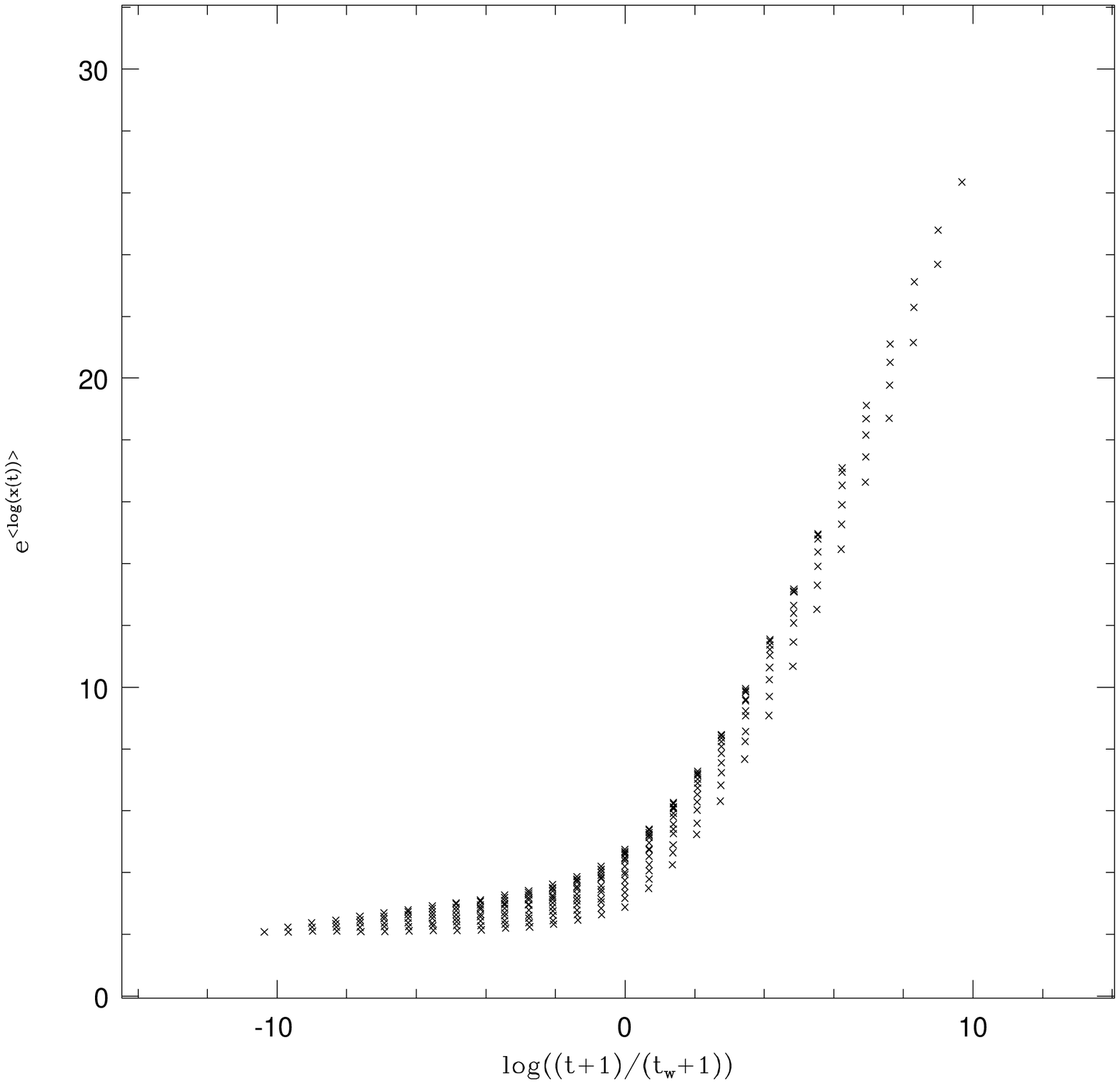}
\caption[a]{A selection of the points of Fig.~1 and 2, for large $t$ and
$t_w$, versus $\ln(\frac{t}{t_w})$.}
\end{figure}

Similar results can be obtained considering the energy-energy
correlation defined by

\be
  C_E(t,t_w) \equiv \overline {(V(x(t))-V(x(t_w)))^2}\ ,
\ee
where $V$ is defined by equation (\ref{E_POT}).

We expect that $C_E(t,t_w) \sim \ln(\frac{t}{t_w})^2$, independently on the
value of $\alpha$.  For the same reason we also expect that the
expectation value of $V$:

\be
  E(t) = \overline {(V(x(t))}
\ee
behaves as $-\ln(t)$ for large $t$, independently on $\alpha$.

Such a scaling law is very well satisfied. In Figures $4-6$ we give
the analogous of Figures $1-3$, and we exhibit the analogous scaling
laws (which in this case are even better than in the former one). In
Fig.~4 we give $-V(t)$ as a function of $\ln(t)$, and we join points
of equal $t_w$. The large $t$ linear behavior is very clear. In
Fig.~5 the same points, but we join point of equal $\frac{t}{t_w}$ ratio.
Again, the curves tend to constant values. In Fig.~6 the rescaling of
the points with $t>32$, $t_w>32$, which again work with very good
accuracy. Here it is very clear, for example, that in the $t<t_w$
asymptotic region the walker does not gain any energy.

\begin{figure}
\epsffile[22 206 565 690]{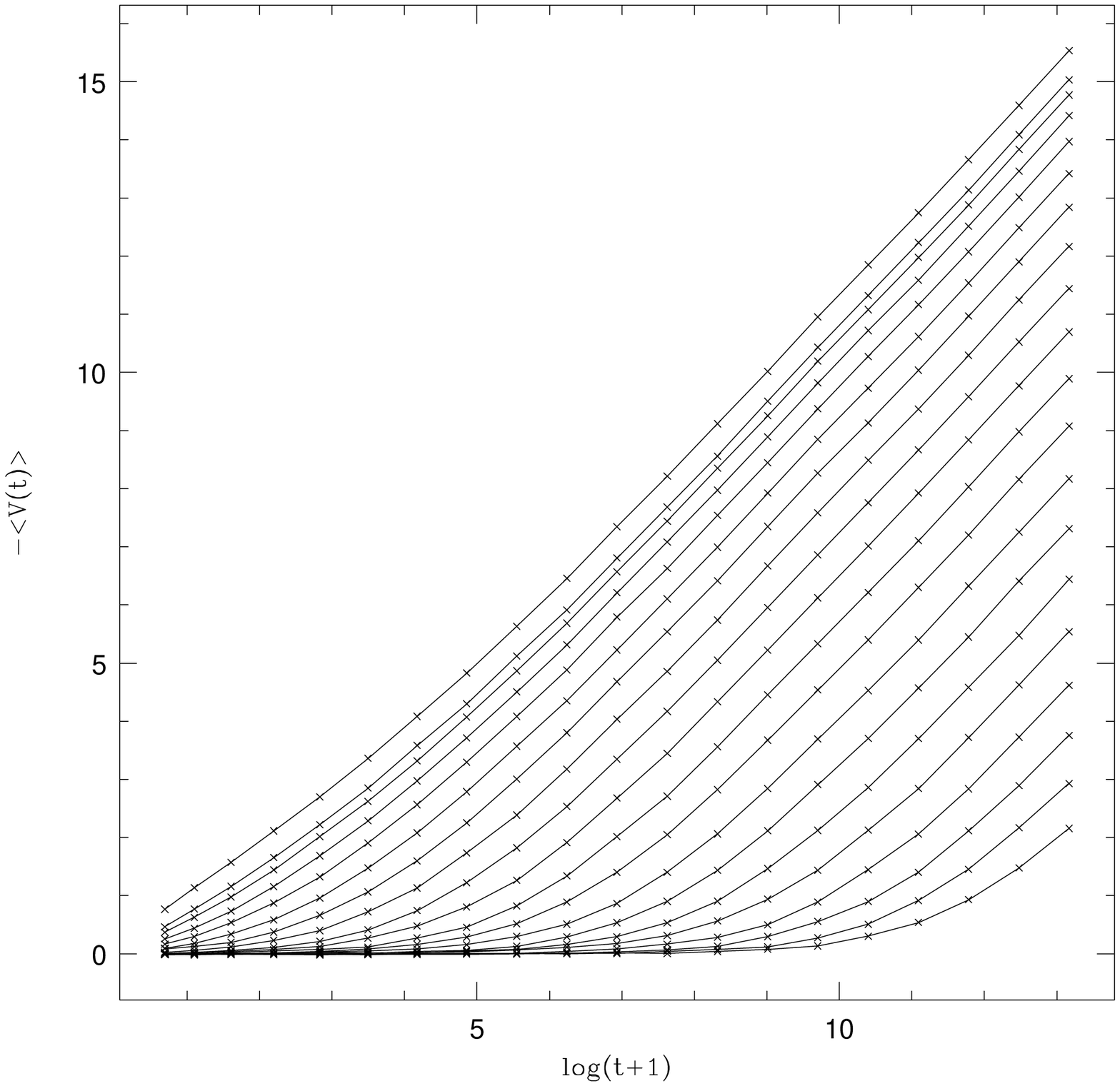}
\caption{As in Fig.~1 (lines join points with constant $t_w$), but for
$-V(t)$ versus $\ln(t)$.}
\end{figure}

\begin{figure}
\epsffile[22 206 565 690]{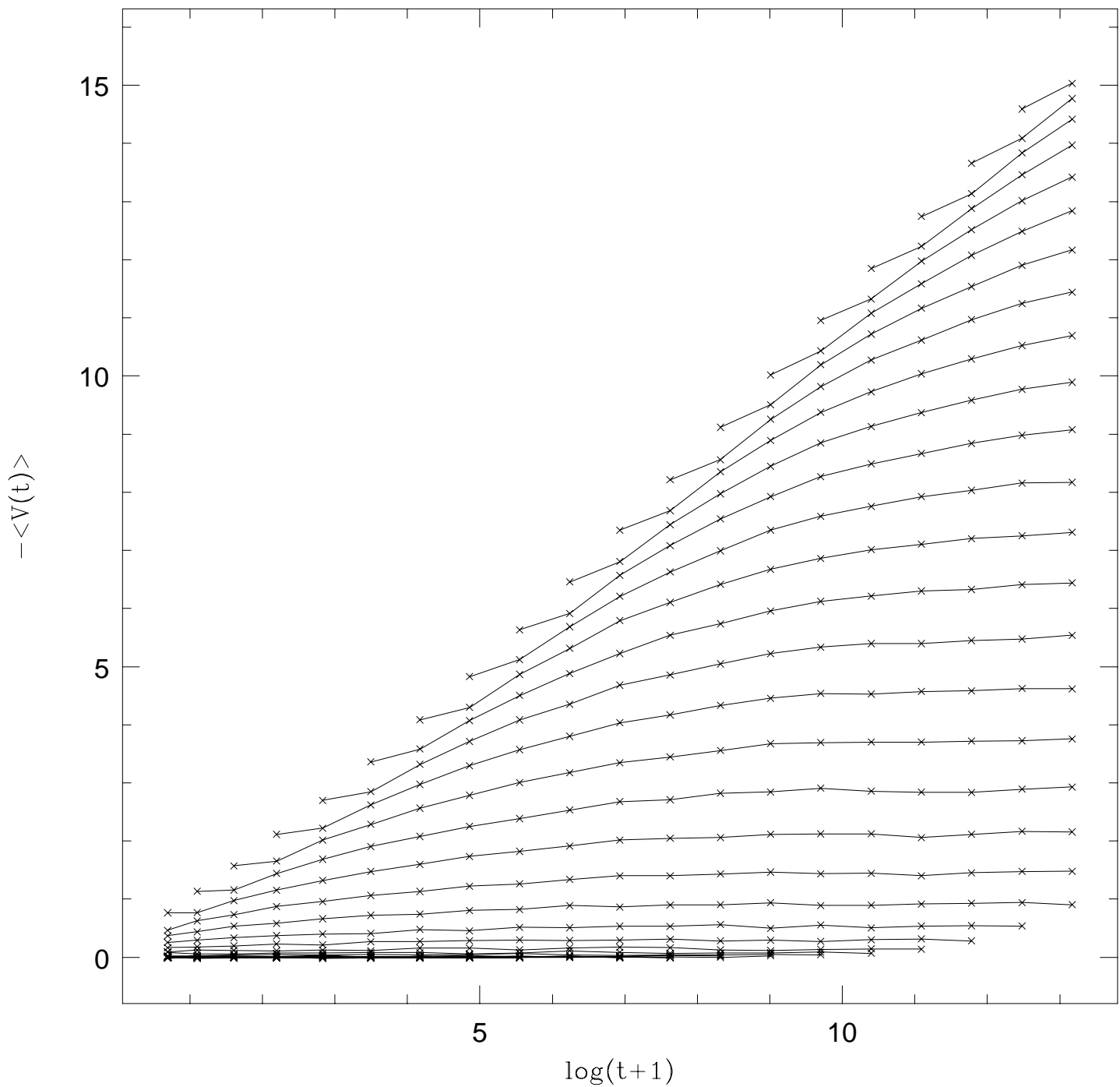}
\caption[a]{As in Fig.~5, but lines join points with constant
$\frac{t}{t_w}$.}
\end{figure}

\begin{figure}
\epsffile[22 206 565 690]{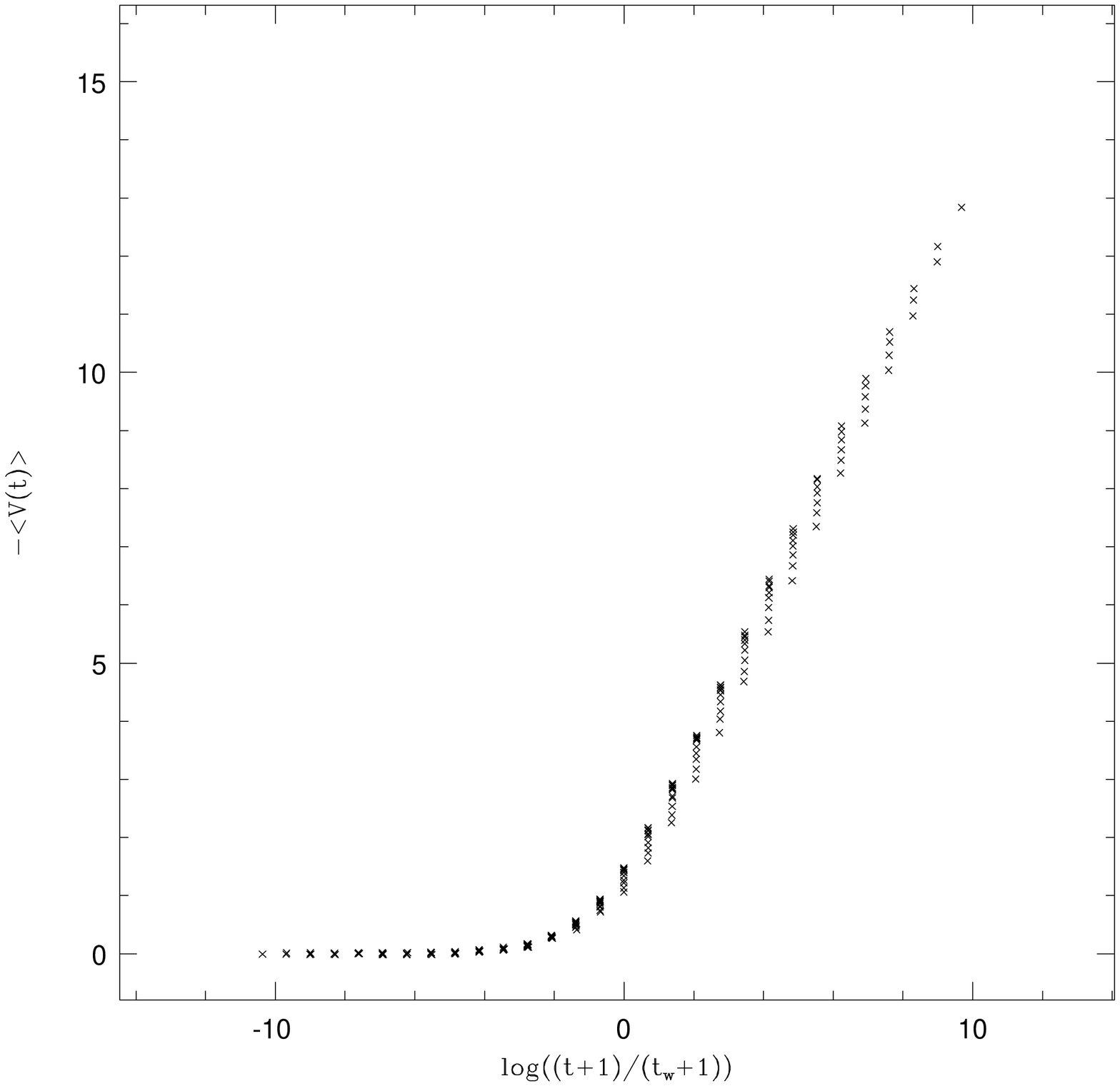}
\caption{As in Fig.~3, but for $-V(t)$.}
\end{figure}

Let us conclude.  The numerical results we have presented here bring
good evidence for the correctness of the simple aging scaling relation

\be
  {\cal C}(t,t_w) \simeq {\cal F}(\frac{t}{t_w})\ ,
\ee
for the correlation function of the logarithm of the distance and of
the energy.

In our model simple aging is correct.  This conclusion seems to us
interesting because of its potential general implications, and as far
as the simplicity of the model is such that a more sound analytic (and
may be rigorous) derivation of these results does not seem impossible.

As a further development of this work, we notice it could be
interesting to study finite volume effects for large, but finite
lattice size (i.e., with reflecting or periodic boundary conditions)
and to study the modification of our results for very large time. The
extension of the model to higher dimensional cases, where the
structure of minima is more complex, will be probably instructive.

\vfill

\end{document}